\documentclass[aps,prb,showpacs,twocolumn,superscriptaddress]{revtex4}

\usepackage{graphicx}
\usepackage{amsmath}
\usepackage{natbib}
\usepackage{bm}
\usepackage[dvips]{color}

\begin{document}

\title{The coordination dependence of electric-field gradients and
hyperfine fields \\ at 5sp impurities on fcc metal surfaces}
\date{\today}

\author{S.~Cottenier}
\email[Electronic address: ]{Stefaan.Cottenier@fys.kuleuven.ac.be}
\affiliation{Instituut voor Kern- en Stralingsfysica, Katholieke Universiteit
Leuven, Celestijnenlaan 200 D, B-3001 Leuven, Belgium}
\author{V.~Bellini}
\email[Electronic address: ]{Bellini.Valerio@unimore.it}
\affiliation{INFM-National Research Center on nanoStructures and bioSystems at
Surfaces (S3) \\ and Dipartimento di Fisica, Universit\`a di Modena e Reggio
Emilia, Via Campi 213/A, I-41100 Modena, Italy}
\author{M.~\c{C}akmak}
\affiliation{Instituut voor Kern- en Stralingsfysica, Katholieke
Universiteit Leuven, Celestijnenlaan 200 D, B-3001 Leuven,
Belgium} \affiliation{Department of Physics, Gazi University,
Teknikokullar, TR-06500 Ankara, Turkey}
\author{F.~Manghi}
\affiliation{INFM-National Research Center on nanoStructures and bioSystems at
Surfaces (S3) \\ and Dipartimento di Fisica, Universit\`a di Modena e Reggio
Emilia, Via Campi 213/A, I-41100 Modena, Italy}
\author{M.~Rots}
\affiliation{Instituut voor Kern- en Stralingsfysica, Katholieke Universiteit
Leuven, Celestijnenlaan 200 D, B-3001 Leuven, Belgium}

\begin{abstract}

We present a comparison between accurate \textit{ab initio} calculations and
a high-quality experimental data set (1986-2002) of electric-field gradients
and magnetic hyperfine fields of Cd at different sites on Ni, Cu, Pd and Ag
surfaces. Experiments found a systematic rule to assign surface sites on
(100) and (111) surfaces based on the main component of the electric-field
gradient, a rule that does not work for (110) surfaces. Our calculations show
that this particular rule is a manifestation of a more general underlying
systematic behavior. When looked upon from this point of view, (100), (111)
\emph{and} (110) surfaces behave in precisely the same way. The
experimentally observed parabolic coordination number dependence of the Cd
magnetic hyperfine field at Ni surfaces is verified as a general trend, but
we demonstrate that individual cases can significantly deviate from it. It is
shown that the hyperfine fields of other 5sp impurities at Ni surfaces have
their own, typical coordination number dependence. A microscopic explanation
for the different dependencies is given in terms of the details of the s-DOS
near the Fermi level.

\end{abstract}
\maketitle


\section{Introduction}
\label{intro}

Nuclear probe techniques, such as M\"{o}ssbauer spectroscopy,
Perturbed Angular Correlation spectroscopy (PAC), Nuclear Magnetic
Resonance (NMR) and others, have proven in the past to be
versatile tools for the study of a broad set of phenomena. By
simultaneously detecting both electric-field gradients (EFG) and
magnetic hyperfine fields (HFF), it has been possible to obtain a
finger print of the electronic and magnetic configuration near and
at the probe nucleus.
\cite{vianden1988,schatz1988,schatz1993,haas1994,voigt1990,bertschat1998}
Ab-initio band structure calculations have assumed an important
role in supporting, improving or confuting the interpretation of
experimental data. Especially for such sensitive properties as
HFF's or EFG's, the availability of these parameter-free
calculations with an accuracy that is comparable with the
experimental error bars, has opened new possibilities. For
instance, the general behavior of the HFF induced on sp- and
d-impurities embedded in bulk magnetic materials, especially in Fe
and Ni, has been well characterized and the calculated values
compared fairly well with the experimental
ones.\cite{akai1984,bluegel1987,akai1990,korhonen2000,cottenier2000}
Similarly, the systematics of the EFG on 5sp-impurities in
non-cubic metals (Zn, Cd, Sb) have been studied by experiments and
calculations (Ref.~\onlinecite{Haas2000} and references therein).
Selected calculations have been performed on more exotic systems.
To name a few, the HFF and EFG on a Cd impurity atom at Fe/Co
\cite{myarticle5}and Fe/Ag \cite{rodriguez2001a,rodriguez2001b}
interfaces have been studied, focussing on the relation between
the HFF induced on the radioactive probe atom and the magnetic
moment profile across the interface. Similar studies have been
performed on magnetic nanoclusters embedded in Ag
\cite{nogueira1999} and Cu \cite{frota-pessoa2001} matrices. In
thin layers of \textit{fcc} Fe on a Cu substrate, HFF and EFG at
the surface and interface have been calculated. \cite{Gomez2002}

If, for a few host materials, the dependence of HFF or EFG on the
atomic number of the impurity atom has been studied in detail,
less investigated is the variation in these properties for one
specific probe atom in different environments derived from the
same host material. An example of such a study -- motivated by the
unique capability of nuclear probe techniques to follow the
diffusion of single atoms -- is provided by the experiments of
Voigt et al., \cite{voigt1990} recently extended and completed by
Potzger et al. \cite{potzger2002} Those authors put a Cd atom at
different low-index Ni-surfaces and at kinks and steps on those
surfaces, and measured the HFF and EFG at Cd. Overlooking the now
fairly complete data set, Potzger et al.\ conclude that the HFF at
Cd depends more or less parabolically on the \emph{number} of
Ni-atoms in the first nearest neighbor shell (NN-shell) (black
dots in Fig.~\ref{fig2}), and not on the \emph{exact position} of
those Ni-neighbors (dubbed `symmetry independence of the HFF' in
Ref.~\onlinecite{Prandolini2003}). This is attributed to a gradual
change in the local DOS, not further specified. One of the two
main goals of the present paper is to assess the validity of this
claim, to elucidate the possible physical mechanism behind it
(Sec.~\ref{results2a}), and to try to extend this `rule' to
impurity atoms other than Cd (Sec.~\ref{results2b}), all this by
calculating the HFF and EFG at Cd in different Ni-environments by
ab~initio methods. A second goal is to examine the generality of
common guidelines for site identification of Cd on (100) and (111)
metal surfaces based on EFG measurements, as were established from
experiments. \cite{schatz1993,haas1994,haas1994b} We will
demonstrate that these guidelines are a manifestation of a simple
and general physical mechanism that is valid for \textit{all}
low-index surfaces (Sec.~\ref{results1}). Our work follows
pioneering cluster calculations by B.~Lindgren et al.\
\cite{Lindgren1987, Lindgren1990,lindgren1993,lindgren2002}, who
investigated HFF and EFG of Cd as adatom or in a terrace site at
(100) and (111) Ni surfaces, and a more systematic work  by
Ph.~Mavropoulos et al.\ \cite{mavropoulos1998} for probe atoms
belonging to the whole 4sp series (Cu to Sr), placed on Ni and Fe
(100) surfaces. Independent from our work, some of the questions
that will be discussed here were  studied very recently by
Ph.~Mavropoulos~\cite{Mavropoulos2003} using a different method
(full-potential KKR), a different exchange-correlation functional
(LDA) and not considering atomic relaxations.


\section{Method and details of the calculations}
\label{method}

For the present calculations we have employed state of the art
first-principles techniques, developed within the Density
Functional Theory (DFT) \cite{hohenberg1964,sham1965}. Most of the
calculations have been performed using the full-potential
augmented plane wave + local orbitals (APW+lo) method as
implemented in the WIEN2k package.\cite{wien2k} We have simulated
the surface+impurity complexes by using the slab-supercell
approach. The impurities have been put both in an adatom site and
at a substitutional position in the surface layer (called the
terrace site from now on) at the three low-Miller-index surfaces,
which in turn were composed of repeated slabs of 5 to 7 substrate
layers (up to 11 for test calculations), separated by some vacuum
space. The size and shape of the cells were chosen such to avoid
artificial interaction between the impurity atoms. For symmetry
reasons, the impurity atoms were placed at both sides of the
slabs. Test calculations showed that about 6 \AA\ of vacuum
(equivalent to about 3.5 atomic layers) is sufficient to decouple
the Cd atoms at the two surfaces, and that a similar distance is
needed to decouple the Ni atoms within the same plane. More
details of the slabs and the vacuum are given in tables later in
the paper. As exchange-correlation functional, the PBE generalized
gradient approximation (GGA) \cite{perdew1996} was used. A few
test calculations (see later) have been performed with the local
density approximation (LDA). In the APW+lo method, wave functions,
charge density, and potential are expanded in spherical harmonics
within nonoverlapping atomic spheres of radius R$_{MT}$ and in
plane waves in the remaining space of the unit cell. R$_{MT}$
values of 2.2 a.u.\ for Ni and 2.4 a.u.\ for the impurity were
chosen throughout. The maximum $\ell$ for the waves inside the
atomic spheres was confined to $\ell_{max}=10$. For the majority
of the cases, the wave functions in the interstitial region were
expanded in plane waves with a cutoff of
K$_{max}$=7.0/R$_{MT}^{min}=3.18$~a.u.$^{-1}$ or better, while the
charge density was Fourier expanded up to
G$_{max}$=16~$\sqrt{\mathrm{Ry}}$. For a few large cells, a lower
K$_{max}$=2.73~a.u.$^{-1}$ was taken, which was tested to yield
still reasonable forces. For the Brillouin zone sampling, a mesh
of special k points corresponding to at least a $14 \times 14
\times 14$ mesh (104 points) for bulk fcc Ni was taken. The
lattice constant of Ni used throughout the calculations is the one
which minimizes the GGA bulk total energy, i.e. $a^{GGA}_{Ni} =
3.51$ \AA. For Cu, Pd and Ag, the experimental lattice constants
were used.

When the perfect crystal structure is broken, as in presence of an impurity,
atomic relaxations of the atoms in the vicinity of impurity appear. Such
relaxations have rather long range character, and lead to a volume change of
the crystal, as testified by refined lattice parameter measurements.
Additionally, since they alter the nature and strength of bonds, they change
local properties such as HFF and EFG as well. Atomic relaxations tend to be
the larger in size the more open is the crystal structure (e.g.\ zincblend)
\cite{settels1999}, but it has been shown in the literature to be important
also in close-packed matrices for HFF's \cite{cottenier2000,korhonen2000} and
even more for EFG's \cite{Petrilli1998}. As taking into account atomic
relaxations is computationally quite expensive, we have speeded up the
calculations by using a combination of methods. First, the atomic positions
were relaxed using the pseudopotential plane wave VASP code
\cite{vasp1,vasp2}. The all-electron APW+lo code was then used in a second
stage to further relax the atoms to their equilibrium positions (with forces
less than 1 mRy/a.u.), and to calculate the hyperfine properties.

Before proceeding with the discussion of the results, it is important to
assess the accuracy of our calculated HFF's or EFG's. To this end, let us
look at a well-known case, Cd in bulk Ni. The experimental value is -6.9 T
(4.2 K) \cite{shirley1968}. Our calculation, carried out for a large $2\times2\times2$ 32-atom
supercell with relaxed nearest neighbors, yields -12.0~T. The calculation
included SO-coupling (in contrast to all other calculations reported in this
paper), from which we could obtain the orbital and dipolar contributions to
the hyperfine field. They were smaller than 0.01 T, however, as could be
expected for a closed shell atom as Cd ([Kr]5s$^2$4d$^{10}$), such that the
so-called Fermi contact term \cite{fermi1930} is the only contribution to the
HFF. The Ni-moment was put in the [111] direction, as in nature. Without
relaxation, the HFF was -9.8~T. Having taken all these precautions, we must
attribute the 5.1~T difference with experiment entirely to the approximations
contained in GGA. This magnitude of error is similar to the 4.2~T
overestimation of the HFF for Cd in Fe, and is of the same magnitude as was
found for all 5$^{\mathrm{th}}$ period impurities in Fe.
\cite{cottenier2000,korhonen2000} This is the state-of-the-art for the
accuracy that can be obtained currently for calculated hyperfine fields. In
what follows, we can therefore expect calculated hyperfine fields that are 5
T too negative if this error is independent on the coordination number, or
alternatively - if the error is more random and does not depend on the
coordination number - we have to put error bars of ±5 T on our results.
Electric-field gradients instead can be calculated with more accuracy, and we
expect here an error bar better than ($\pm 2 \cdot 10^{21}$ V/m$^2$) (see for
instance Ref.~\onlinecite{blaha1995}).


\section{Electric-Field Gradients}
\label{results1}

For an efficient discussion later in this section, we first
summarize a few properties and conventions with respect to the
electric-field gradient (EFG) at a nucleus in a crystal (for a
more elaborate discussion, see for instance
Refs.~\onlinecite{Klas1986} and~\onlinecite{Hunger1990b}). The EFG
is a traceless, symmetric tensor of rank 2 (5 components), and is
a measure of the deviation from spherical symmetry of the
electronic charge density around the nucleus under consideration.
In its principal axis system (PAS), the EFG tensor is diagonal
with diagonal elements $V_{\mathrm{xx}}$, $V_{\mathrm{yy}}$ and
$V_{\mathrm{zz}}$ and $V_{\mathrm{xx}} + V_{\mathrm{yy}} +
V_{\mathrm{zz}} = 0$ (traceless, hence specifying 2 out of the 3
quantities is enough). The axes of the PAS are conventionally
labelled such that $\left| V_{\mathrm{zz}} \right| \ge \left|
V_{\mathrm{yy}} \right| \ge \left| V_{\mathrm{xx}} \right|$.
Instead of specifying 2 of the 3 diagonal elements, usually the
couple $\left(V_{\mathrm{zz}}, \, \eta \right)$ is specified, with
$\eta$ the so-called asymmetry parameter which equals
$(V_{\mathrm{xx}} - V_{\mathrm{yy}})/V_{\mathrm{zz}}$ ($0 \le \eta
\le 1$). Often one refers with the term `electric-field gradient'
to V$_{\mathrm{zz}}$ only. Whenever we use the word `EFG', we mean
the tensor with its 5 components. If we mean the main component in
the PAS, we write V$_{\mathrm{zz}}$.

Over the past two decades, a large body of experimental results
has been obtained for the EFG's of Cd at different surfaces of Ni,
Cu, Pd and Ag (see Tab.~\ref{tab0}
\begin{table}
 \begin{center}
 \caption{Experimental absolute values of the principal component
   $\left| V_{\mathrm{zz}} \right|$ of the electric-field gradient tensor of Cd
   at various metallic surfaces, with references tracked down as much as
   possible to the original sources. Units: $10^{21}$~V/m$^2$. The lowest
   available measurement temperatures were chosen, and are specified below
  (0~K if a reliable extrapolation was available). \label{tab0}}
  \begin{ruledtabular}

   \begin{tabular}{lcccc}
                    &   Ni      &   Cu      &   Pd      &   Ag      \\ \hline
   (100) terrace    &  8.2  \footnote{unpublished Ref.~\onlinecite{thesis-voigt}, mentioned in Refs.~\onlinecite{lindgren1993} and~\onlinecite{bertschat1998}} \footnote{Ref.~\onlinecite{bertschat1998}, T=270-300~K}    &   10.3 \footnote{Ref.~\onlinecite{Klas1986}, T=0~K}     &   8.2  \footnote{Ref.~\onlinecite{Fink1993}, $\eta=0.16$ (!), T=0~K}   &   7.5 \footnote{Ref.~\onlinecite{Wesche1989}, T=0~K}    \\
   (100) adatom     &  2.8 \footnote{unpublished Ref.~\onlinecite{thesis-voigt}, mentioned in Ref.~\onlinecite{lindgren1993}} / 0.3 \footnote{unpublished Ref.~\onlinecite{thesis-voigt}, mentioned in Ref.~\onlinecite{potzger2002}} &   0.8 \footnote{Mentioned in Ref.~\onlinecite{Lindgren1990}: private communication from G.~Schatz}    &   2.8  \footnote{Ref.~\onlinecite{Fink1993}, T=77~K}   &   0.3  \footnote{Ref.~\onlinecite{fink1990},T=77~K}   \\ \hline

   (110) terrace    & --    &   7.9 \footnote{Ref.~\onlinecite{Klas1989}, $\eta=0.74$, T=0~K} &   7.9 \footnote{unpublished Ref.~\onlinecite{Fillebock1997}, mentioned in Ref.~\onlinecite{lindgren2002}. $\eta= 0.97$}  &   7.0 \footnote{Ref.~\onlinecite{Wesche1989}, $\eta=0.80$, T=77~K} \\
   (110) adatom     & --    &   --  &   8.5 \footnote{unpublished Ref.~\onlinecite{Fillebock1997}, mentioned in Ref.~\onlinecite{lindgren2002}. $\eta= 0.42$}  &   --  \\ \hline

   (111) terrace    & 11.5 \footnote{Ref.~\onlinecite{voigt1990}, T=0~K} / 12.3 \footnote{Ref.~\onlinecite{potzger2002}, T=340~K}   &   10.2 \footnote{Refs.~\onlinecite{Klas1988} and~\onlinecite{Klas1989}, T=0~K}    &   10.2 \footnote{Ref.~\onlinecite{Hunger1990a} and~\onlinecite{Hunger1990b}, T=77~K}  &
   8.6 \footnote{Ref.~\onlinecite{Wesche1989}, T=0~K} \\
   (111) adatom     &   1.0 \footnote{Ref.~\onlinecite{potzger2002}, T=36~K} & --  &   0.4 \footnote{Ref.~\onlinecite{Hunger1990a} and~\onlinecite{Hunger1990b}, T=77~K}
   & --

    \end{tabular}
  \end{ruledtabular}
 \end{center}
\end{table}
for a summary of the experimental work, with references tracked
down as much as possible to the original sources). In contrast to
what the structure of Tab.~\ref{tab0} might suggest, the
assignment of the experimental values to either terrace or adatom
sites does not happen in a direct way. For (100) and (111)
surfaces, it is easy to separate terrace and adatom sites from all
other ones (steps, kinks) due to the axial symmetry ($\eta = 0$)
of the EFG and the perpendicular orientation of the z-axis of its
PAS. It is much harder to deduce which of both is the terrace site
and which one is the adatom site. Only for a few cases a thorough
experimental analysis has been performed: Ag(100) \cite{fink1990}
and Pd(111) \cite{Hunger1990a, Hunger1990b}. Arguments include the
thermal annealing behavior, the fraction of different sites
dependent on the type of vicinal cut, and not at least a
comparison with early ab~initio cluster calculations.
\cite{Lindgren1987, Lindgren1990} Apart from these two well-tested
cases, the partial information from thermal annealing behavior
only (Cu(111) \cite{Klas1989}) or from agreement with cluster
calculations only (Cu(100) \cite{Klas1989}) further supports the
assignment in Tab.~\ref{tab0}. To our knowledge, the results on
Pd(110) are published only recently and in an indirect way
(unpublished Ref.~\onlinecite{Fillebock1997}, mentioned in
Ref.~\onlinecite{lindgren2002}). Prior to this information,
Tab.~\ref{tab0} was commonly summarized by the following practical
`rule': \emph{on a metal surface, a large $V_{\mathrm{zz}}$ for Cd
indicates the terrace site, while a low $V_{\mathrm{zz}}$
indicates the adatom site}. This rule has been applied e.g.\ by
Potzger et al.\ \cite{potzger2002} to assign the adatom and
terrace sites on Ni(100) and Ni(111). If the rule would be not
universal, this site assignment could be wrong, and consequently
also the conclusion on the parabolic coordination number
dependence of the Cd-HFF (Fig.~\ref{fig2} and
Section~\ref{results2a}). Therefore, we will first examine the
validity of this rule. Is it really valid for all (100) and (111)
surfaces of the nonmagnetic metals Cu, Pd and Ag (i.e.\ also for
the cases that are not experimentally tested in detail)? Can the
rule be transferred without changes to the magnetic metal Ni? And
is it really true only for (100) and (111) surfaces, and not for
(110) (as the recent Pd information suggests)? Finally, what is
the physical mechanism that lies behind this rule?

In response to these questions, we present in Tab.~\ref{tab1}(a)
\begin{table}
 \begin{center}
  \caption{(a) Column 1: unrelaxed slab calculations for $V_{\mathrm{zz}}$ of Cd on Ni surfaces. Column 2: idem, but relaxed.
  Column 3: unrelaxed cluster calculations for $V_{\mathrm{zz}}$ of Cd on Ni surfaces, when available. If no values were
  available, results for Cu or Pd surfaces are given instead. Column 4: Absolute value of the experimental $V_{\mathrm{zz}}$ for Cd
  on Ni surfaces, if available (if not, values for a Pd surface are given). (b) Relaxed slab calculations for $V_{\mathrm{zz}}$
  of Cd in terrace and adatom sites at one magnetic and three nonmagnetic (100) surfaces (sign included), compared
  with the absolute value of the corresponding experimental $V_{\mathrm{zz}}$. Units: $10^{21}$~V/m$^2$ everywhere. \label{tab1}}

  \vspace{2mm}
  \centerline{(a)}
  \vspace{2mm}

  \begin{ruledtabular}
   \begin{tabular}{lcccc}
                  &    \multicolumn{2}{c}{APW+lo}    &          Cluster Calc.          &   $\left| \mathrm{Exp.} \right|$     \\
                  &   Unrelax.      &    Relax.      & Ref.[\onlinecite{lindgren2002}] &       (see Tab.~\ref{tab0})     \\\hline
    (100) Terrace &    10.5         &     8.6        &             11.0                &    8.2     \\
    (100) Adatom  &     4.0         &     1.9        &             (--0.8)$_{\mathrm{Cu}}$             &  2.8/0.3   \\
    (110) Terrace &     9.5         &    --9.7        &             (11.2)$_{\mathrm{Pd}}$              &   (7.9)$_{\mathrm{Pd}}$    \\
    (110) Adatom  &    --9.3         &    -6.4        &              (--8.3)$_{\mathrm{Pd}}$                &   (8.5)$_{\mathrm{Pd}}$    \\
    (111) Terrace &    10.8         &    13.0        &             16.5                &  11.5/12.3 \\
    (111) Adatom  &    --1.4         &    --2.9        &             (--6.2)$_{\mathrm{Cu}}$              &    1.0  \\ \hline
    (110) Terrace &     $\eta = 0.83$    &    $\eta = 0.74$    &             $\eta_{\mathrm{Pd}} = 0.97$             &     $\eta_{\mathrm{Pd}} = 0.97$    \\
    (110) Adatom  &     $\eta = 0.18$    &    $\eta = 0.23$    &             $\eta_{\mathrm{Pd}} = 0.28$             &     $\eta_{\mathrm{Pd}} = 0.42$ \\
   \end{tabular}
  \end{ruledtabular}

  \vspace{2mm}
  \centerline{(b)}
  \vspace{2mm}

  \begin{ruledtabular}
   \begin{tabular}{ccccc}
          &\multicolumn{2}{c}{(100) terrace} & \multicolumn{2}{c}{(100) adatom} \\
          &     APW+lo      &     $\left| \mathrm{Exp.} \right|$       &     APW+lo      &      $\left| \mathrm{Exp.} \right|$      \\ \hline
    Ni    &      8.6        &     8.2        &      1.9        &    2.8/0.3     \\
    Cu    &      9.9        &    10.3        &    --2.0        &      0.8       \\
    Pd    &      9.1        &     8.2        &      3.4        &      2.8       \\
    Ag    &      8.5        &     7.5        &      0.4        &      0.3       \\
   \end{tabular}
  \end{ruledtabular}

 \end{center}
\end{table}
the calculated $V_{\mathrm{zz}}$ for both relaxed and unrelaxed
structures, together with their assigned experimental values
(taken from Tab.~\ref{tab0}). We did not find in the literature
any experimental investigation for Cd at the Ni(110) surface.
Instead, for a qualitative comparison, the data for Cd at the
Pd(110) surface are enclosed in parenthesis in Tab. \ref{tab1}(a).
We also include the results obtained with the cluster method by
B.~Lindgren. \cite{lindgren2002} Unfortunately, the cluster
calculations of Lindgren spanned all the orientations only for the
Cu surface, while for Cd on Ni only the terrace sites at Ni(100)
and (111) surfaces have been simulated. Results for Cu and Pd
surfaces are therefore included in Tab.~\ref{tab1}(a) in
parentheses as well. We recall here that differently to the
cluster results of Lindgren, our calculations are for bulk slabs
and not for non-periodic clusters. Moreover we do a `full'
relaxation of all the atomic positions in the cell, while due to
insufficient accuracy of the method, no real forces have been
calculated by Lindgren: the distance between Cd and the surface
was fixed at the experimental distance, and the surrounding Ni
atoms were kept in their ideal \textit{fcc} positions (in contrast
to this, the Cd atom in our \emph{un}relaxed calculations was at
an ideal \textit{fcc} lattice site). Our fully relaxed results
compare generally very well with the experiments, predicting for
(100) and (111) a low $V_{\mathrm{zz}}$ (close to zero) for all
adatom sites, and a considerably higher $V_{\mathrm{zz}}$ for the
terrace site, in compliance with the experimental rule. For the
(110) surface, terrace and adatom sites have a comparable
$V_{\mathrm{zz}}$, again in agreement with the experimental
observation. Taking into account the rather different
approximations, there is also reasonable agreement between our
unrelaxed slab calculations and the unrelaxed cluster results, the
former being generally closer to experiment (especially for
(111)). For the (110)-surface, the EFG is not axially symmetric
($\eta \neq 0$). Values for $\eta$ are included in
Tab.~\ref{tab1}(a) as well. Relaxation can be seen to have a
moderate influence on $\eta$, and the considerably larger $\eta$
in terrace sites as seen in experiments, is well-reproduced.
Direct comparison with experimental values for $\eta$ is difficult
with the present data set, however (no experiments for Ni(110) and
no relaxed calculations for Pd(110)).

For the (110) terrace site we observe a change in the sign of
$V_{\mathrm{zz}}$, when moving from the unrelaxed to the relaxed
system. This is readily explained as follows. $V_{\mathrm{zz}}$ is
defined as the largest component of the electric field gradient
tensor in its principal axis system. But it might happen that
there are two components of similar size, that have then
necessarily opposite signs and yield a large asymmetry parameter
$\eta$ (=close to 1). This is exactly what we found for the (110)
terrace case. There is a large positive component along the normal
on the (110) surface, while an almost as large negative component
exists along the [1${\bar 1}$0] direction in the (110) plane. Upon
relaxation the latter negative component becomes in absolute value
larger than the positive one, and the latter axis becomes the one
along which $V_{\mathrm{zz}}$ is defined. In the adatom case, the
z-axis is the [001] direction in the (110) plane. Experiments on
the Ni(110) surface predict instead the EFG z-axis to lay along
the normal to the (110) plane. This apparent deviation between
experiment and calculation is only a consequence from the way how
$V_{\mathrm{zz}}$ is defined, not from a real discrepancy. For the
other two substrate orientations, (110) and (111), the asymmetry
parameter $\eta$ is zero and there is therefore no doubt on which
component is the largest one: we find in all cases the principal
axis along the surface normal, as in experiment.

Having verified that the practical site assignment rule is valid
for magnetic Ni(100) and Ni(111) surfaces, and not for Ni(110), we
now examine for one type of surface -- (100) -- whether other
elemental metal substrates show the same behavior.
Tab.~\ref{tab1}(b) shows the calculated and experimental
$V_{\mathrm{zz}}$ at the (100) surfaces of Ni, Cu, Pd and Ag. The
calculations are in very satisfying agreement with the
experiments, and it is safe now to conclude that the EFG in all of
these 24 surface sites behaves in the same way.

Finally, we search for the physical mechanism that is responsible
for this systematic behavior. Instead of considering for this the
Ni-calculations from Tab.~\ref{tab1}, we calculated the EFG of Cd
at a complete set of unrelaxed Pd-surfaces. This avoids the
complication of an EFG coming from spin up and spin down
electrons, and it allows for a more direct comparison with
experiment because for Pd all six cases are measured
(Tab.~\ref{tab0}). It does not harm not to consider relaxation
here, because now we are looking for a global mechanism, not for
fine details. The results are given for every diagonal component
of the EFG separately in Tab.~\ref{tab2},
\begin{table*}
 \begin{center}
  \caption{The three calculated ($V_{\mathrm{ii}}^{\mathrm{calc}}$, unrelaxed) and experimental
  ($V_{\mathrm{ii}}^{\mathrm{exp}}$, see Tab.~\ref{tab0} for references) components of the EFG tensor in its
  PAS for Cd at 6 Pd surface sites. Units: $10^{21}$~V/m$^2$. The usual convention for labelling the axes of the PAS is \emph{not} followed:
  the z-axis is perpendicular to the surface, x- and y-axes are in the surface plane. For (100) and (111) surfaces, all directions
  in the surface plane are equivalent for the EFG. For the (110) surface, our x-axis refers to the [1${\bar 1}$0] direction, and our y-axis to the
  [001] direction. The principal component -- which would be labelled $V_{\mathrm{zz}}$ in the usual convention --
  is printed in bold. The sign of the experimental values was not measured, but is chosen here to agree with the calculations
  (for the small numbers for NN=3, this is nothing more than a guess). The number of Cd-5p electrons split into
  $n_{\mathrm{p_x}}^{\mathrm{calc}}$, $n_{\mathrm{p_y}}^{\mathrm{calc}}$ and $n_{\mathrm{p_z}}^{\mathrm{calc}}$ is given as well. \label{tab2}}
  \begin{ruledtabular}
   \begin{tabular}{l|c|ccc|ccc|ccc}
     & $\sharp$ NN & $V_{\mathrm{xx}}^{\mathrm{calc}}$ &
    $V_{\mathrm{yy}}^{\mathrm{calc}}$& $V_{\mathrm{zz}}^{\mathrm{calc}}$ & $V_{\mathrm{xx}}^{\mathrm{exp}}$ & $V_{\mathrm{yy}}^{\mathrm{exp}}$ & $V_{\mathrm{zz}}^{\mathrm{exp}}$ & $n_{\mathrm{p_x}}^{\mathrm{calc}}$ &
    $n_{\mathrm{p_y}}^{\mathrm{calc}}$ & $n_{\mathrm{p_z}}^{\mathrm{calc}}$  \\ \hline

    (111) terrace & 9 & -4.8 & -4.8 & \textbf{9.6} & -5.1 & -5.1 &
    \textbf{10.2} & 0.1478 & 0.1478 & 0.0952 \\

    (100) terrace & 8 & -4.4 & -4.4 & \textbf{8.8} & -4.1 & -4.1 &
    \textbf{8.2} & 0.1321 & 0.1321 & 0.0876 \\

    (110) terrace & 7 & -7.6 & -0.5 & \textbf{8.1} & -7.8 & -0.1 &
    \textbf{7.9} & 0.1214 & 0.0986 & 0.0736 \\

    (110) adatom & 5 & 2.0 & \textbf{-6.2} & 4.2 & -2.5 & -6.0 &
    \textbf{8.5} & 0.0681 & 0.0972 & 0.0651 \\

    (100) adatom & 4 & -1.4 & -1.4 & \textbf{2.8} & -1.4 & -1.4 &
    \textbf{2.8} & 0.0645 & 0.0645 & 0.0580 \\

    (111) adatom & 3 & 0.4 & 0.4 & \textbf{-0.8} & 0.2 & 0.2 & \textbf{-0.4} & 0.0424 & 0.0424
    & 0.0574

   \end{tabular}
  \end{ruledtabular}
 \end{center}
\end{table*}
together with the corresponding experimental components (mind the
unconventional labelling of the axes, as specified in the caption
of Tab.~\ref{tab2}). Considering the fact that these are unrelaxed
calculations, they reflect the experimental trend very well and
hence these Pd-surfaces form a reliable model system to search for
the physical mechanism.

\begin{figure*}
 \begin{center}
  \includegraphics[width=18cm,angle=0]{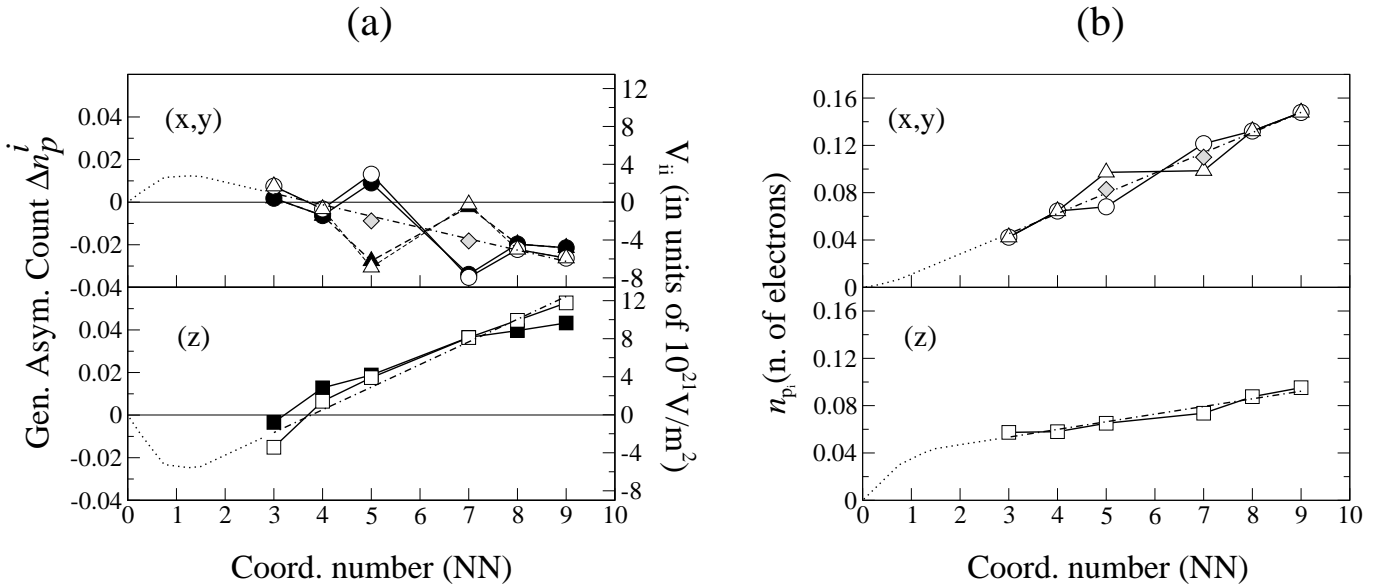}
  \caption{(a) Black symbols: calculated components of the EFG (see Tab.~\ref{tab2}, black symbols refer to
  the right scale). White symbols: generalized asymmetry counts (see text), calculated from the $n_{p_i}$ in
  Tab.~\ref{tab2} (white symbols refer to the left scale). Gray symbols (for (110) surfaces only): average of
  $\Delta n^x_p$ and $\Delta n^y_p$, which would be the value of $\Delta n^x_p$=$\Delta n^y_p$ in the hypothetical
  case of axial symmetry ($\eta=0$). Circles, triangles and squares are relative to the x,y,z components
  respectively of the plotted quantities. Dotted-dashed lines: $\Delta n^i_p$ obtained from the least
  squares fit in Fig.~\ref{fig0}(b). The dotted lines are calculated using the extrapolation of Fig.~\ref{fig0}(b).
  (b) The $n_{p_i}$ from Tab.~\ref{tab2}. Gray symbols (for (110) surfaces only): average of $n_{p_x}$ and $n_{p_y}$
  in the hypothetical case of axial symmetry ($\eta=0$). Dotted-dashed lines: least squares fit, with tentative
  extrapolation (dotted line) to NN=0 (free atom). Both in (a) and (b) the data for the xy-plane and for the
  z-direction are displayed separately for clarity. \label{fig0}}
 \end{center}
\end{figure*}

Already in the very first successful \textit{ab initio}
calculations of EFG's in metals, it has been shown that magnitude
and sign of the principal component $V_{\mathrm{zz}}$ are
reflected in the so-called asymmetry count $\Delta n_p$ of the
p-electrons (for Cd, the completely filled d-shell does not
contribute to the EFG). The asymmetry count is defined as $\Delta
n_p \, = \, \frac{1}{2} \left(n_{p_x} + n_{p_y} \right) -
n_{p_z}$, with $n_{p_i}$ the number of electrons in the
$p_i$-orbital (see Ref.~\onlinecite{blaha1988} for the first
analysis, and
e.g.~Refs.~\onlinecite{Ambrosch1989,Schwarz1990,Lany2000,Cottenier2002}
for applications): spherical symmetry $\left( n_{p_x} = n_{p_y} =
n_{p_z} \right)$ leads to $\Delta n_p = 0$ (hence
$V_{\mathrm{zz}}=0$), charge accumulation along the z-axis
($n_{p_x}$ large) leads to $\Delta n_p < 0$ (hence
$V_{\mathrm{zz}} < 0$), etc. For our analysis, we slightly extend
this idea and take as a working hypothesis that sign and magnitude
of \textit{any} diagonal component $V_{\mathrm{ii}}$ is reflected
in a \textit{generalized} asymmetry count $\Delta n_p^i \, = \,
\frac{1}{2} \left(n_{p_j} + n_{p_k} \right) - n_{p_i}$ (and cyclic
permutations). This hypothesis is checked by Fig.~\ref{fig0}(a),
where the calculated $V_{\mathrm{ii}}$ from Tab.~\ref{tab2} (right
scale) are compared with the asymmetry counts $\Delta n_p^i$
derived from the calculated number of p$_i$ electrons in
Tab.~\ref{tab2} (left scale). With the right choice of scales in
Fig.~\ref{fig0}(a), the $\Delta n_p^i$ and  $V_{\mathrm{ii}}$ data
points almost coincide, which means that apart from a unique
scaling factor both quantities reflect basically the same
behavior. This proves our working hypothesis: the observed
behavior of the components of the EFG is nothing more than a
manifestation of the properties of the distribution of the Cd
5p-electrons over the x-, y- and z-directions. Hence, if we
understand the behavior of the $n_{p_i}$ in Tab.~\ref{tab2}, we
understand the EFG. This greatly facilitates our search, as -- in
contrast to the $V_{\mathrm{ii}}$ -- the $n_{p_i}$ have a direct
and intuitively understandable relation to the chemical bonds
between Cd and its Pd neighbors.

The coordination dependence of the $n_{p_i}$ is visualized in
Fig.~\ref{fig0}(b), together with a least squares fit. The picture
is complicated a bit by the fact that $n_{p_x}$ and $n_{p_y}$ are
not identical to each other for (110) surfaces. To make the
discussion more transparent, we will therefore first assume a
hypothetical (110) surface with axial symmetry ($\eta = 0$) for
which $n_{p_x} = n_{p_y}$ (gray symbols in Fig.~\ref{fig0}, they
are the average of the calculated $n_{p_x}$ and $n_{p_y}$). We now
clearly see an almost linear dependence of $n_{p_x}$ (or
$n_{p_y}$) and $n_{p_z}$ on NN (least squares fit in
Fig.~\ref{fig0}(b)): both $n_{p_x}$ ($n_{p_y}$) and $n_{p_z}$
decrease linearly while reducing the coordination, the decrease
being stronger for $n_{p_x}$ ($n_{p_y}$) than for $n_{p_z}$. If we
calculate the $\Delta n^i_p$ corresponding to the least squares
fits in Fig.~\ref{fig0}(b), we obtain the straight lines in
Fig.~\ref{fig0}(a) (for overall axial symmetry, gray symbols for
NN=(5,7)): this proves that the two-slope model from
Fig.~\ref{fig0}(b) is indeed the key responsible for the observed
behavior of the EFG.

\begin{figure}
 \begin{center}
  \includegraphics[width=6.3cm,angle=270]{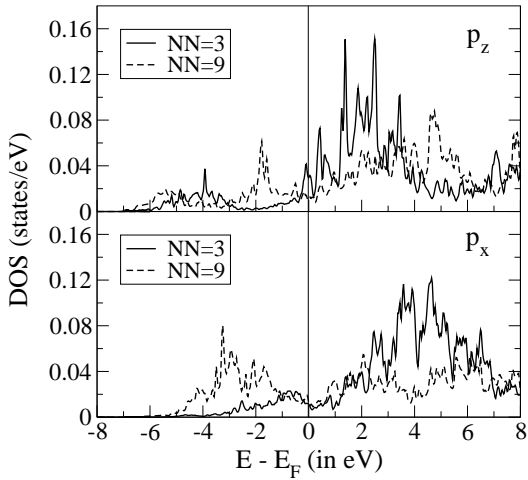}
  \caption{Partial $p_z$- and $p_x$-DOS for unrelaxed Cd on Pd (NN=3 and NN=9). \label{fig1}}
 \end{center}
\end{figure}

Can we understand the two slopes in Fig.~\ref{fig0}(b)? Obviously, $n_{p_i}$
should decrease with decreasing NN: Cd has no native 5p-electrons, hence in
the free atom limit (NN=0) all $n_{p_i}$ should be zero. The reason why
$n_{p_x}$ ($n_{p_y}$) grows faster with NN than $n_{p_z}$ can be understood
from looking at the partial $p_x$- and $p_z$-DOS for two extreme cases (NN=3
and NN=9, see Fig.~\ref{fig1}).
The $p_z$-DOS has two distinct contributions below
E$_{\mathrm{F}}$: a peak about 5 eV below E$_{\mathrm{F}}$ which
nicely correlates with the s-DOS of Pd neighbors \emph{underneath}
the Cd atom, and hence indicates a vertical Pd-Cd s-p
hybridization, and a second peak in the interval $\left[-2, \, 0
\right]$ eV that corresponds with the Pd d-DOS (Pd-Cd d-p
hybridization). This second peak has almost disappeared for NN=3,
which indicates that it originates mainly from Pd-neighbors in the
horizontal plane. For the $p_x$-DOS, there is only one peak,
correlated mainly with the Pd d-DOS of neighbors in the horizontal
plane. When going from NN=9 to NN=3, the number of Pd neighbors in
the horizontal plane quickly drops (6 for NN=9, 4 for NN=8, 2 for
NN=7 and 0 for lower NN), while the number of Pd neighbors
underneath Cd stays rather constant (3, 4 or 5, with the largest
value for intermediate NN). Hence, $n_{p_x}$ should quickly drop
while $n_{p_z}$ tends to stay constant, which is exactly what we
observed in Fig.~\ref{fig0}(b). Reintroducing $\eta \neq 0$ in our
discussion only redistributes the charges within the horizontal
plane, and therefore does not affect our conclusion.

We emphasize that in contrast to the claimed symmetry independence of the HFF
(to be discussed in Sec.~\ref{results2a}), the coordination dependence of the
p-distribution for sure depends on the exact spatial arrangements of the Pd
neighbors. The behavior is so smooth only because we are dealing with fairly
`similar' environments (Pd underneath, vacuum above). Furthermore, a
tentative extrapolation of Fig.~\ref{fig0}(b) to very low coordination sites
(NN=1, 2) suggests a negative $V_{\mathrm{zz}}$ there. This is confirmed for
the NN=2 environment for Cd on Ni to be discussed in Tab.~\ref{tab2}, where
we calculate $V_{\mathrm{zz}} = -2.4 \cdot 10^{21}~\textrm{V}/\textrm{m}^2$.

In contrast to the discrete behavior suggested by the practical rule
(`terrace site means large $V_{\mathrm{zz}}$, adatom site means low
$V_{\mathrm{zz}}$'), we conclude that there is a continuous evolution as a
function of the coordination number. The (100) and (111) terrace sites are
both highly coordinated, and therefore have indeed the largest
$V_{\mathrm{zz}}$, while the opposite is true for the (100) and (111) adatom
sites. The (110) terrace and adatom sites have intermediate coordination --
and therefore comparable $V_{\mathrm{zz}}$'s -- which is why the `rule' does
not seem to work for (110). However, there is nothing exceptional about the
(110) surface.


\section{Magnetic Hyperfine fields}
\label{results2}

We now limit ourselves to Ni surfaces again and discuss the results obtained
for the hyperfine fields. Two main questions will be addressed. First, we
want to verify and understand the experimentally proposed parabolic
dependence of the Cd HFF on the number of Ni neighbors in the first
coordination shell (NN), and we examine whether and how the spatial
arrangement of these neighbors influences the HFF. As a second problem, we
broaden the scope to the entire 5sp series, placing 5sp impurity atoms only
in NN=4 (adatom) and NN=8 (terrace) coordinated sites at the Ni(100) surface.
This will allow us to generalize the behavior observed for Cd to other probe
atoms, and to propose qualitative explanations on the mechanisms ruling the
observed HFF's.

\subsubsection{Cd HFF's on Ni surfaces}
\label{results2a}

In this section we focus on Cd probe atoms, placed in terrace and
adatom positions at the three low-Miller-index surface of
\textit{fcc} Ni. This gives access to 6 differently coordinated
sites, in addition to the fully coordinated substitutional bulk
site and to a more artificial bridge site with NN=2. In order to
test the sensibility of the HFF on the details of the cells, we
performed also several calculations by varying the cell size and
number of Ni layers in the slab. All the calculated Cd HFF for the
differently coordinated systems are summarized in Table
\ref{tab3}.
\begin{table*}
 \begin{center}
  \caption{Hyperfine fields (T) of Cd in different Ni-environments. Experimental values are taken
  from Ref.~\onlinecite{potzger2002} \label{tab3}}
  \begin{ruledtabular}
   \begin{tabular}{lccccc}
    &&&\multicolumn{3}{c}{Hyperfine fields (in T)} \\
                  &$\sharp$ NN&           Type of cell          &   non-relaxed   &   relaxed       & Exp \\\hline
        Bulk      &   12      &      ($2\times2\times2$)        & \textbf{--9.8}  & \textbf{--12.0} &--6.9\\
    (100) Bulk    &           &      ($2\times2$ , 7L)          &     --10.0      &     --11.7      &     \\\hline
    (111) Terrace &    9      &    ($\sqrt{2}\times\sqrt{2}$ , 7L)    &     \textbf{--7.4} (12.18 \AA) & \textbf{--13.7} (10.49 \AA) &--6.6\\\hline
    (100) Terrace &    8      &      ($2\times2$ , 5L)          &  --9.8 (7.02 \AA)& --1.9 (5.80 \AA) &--3.5\\
                  &           &      ($2\times2$ , 7L)          & \textbf{--8.3} (7.02 \AA)&   \textbf{--6.7} (5.82 \AA)         &     \\
                  &           & ($\sqrt{2}\times\sqrt{2}$ , 7L) &--3.6 (13.19 \AA)&--5.9 (11.88 \AA)&     \\
     bulk-like (`random') & & ($2\times2\times2$) & --9.9 & --- &  \\
     bulk-like (terrace) & & ($2\times2\times2$) & --8.2 & --- &  \\
     \textit{bcc}-bulk & & ($2\times2\times2$) & --8.9 & --- & \\ \hline
    (110) Terrace &    7      &   ($2\times\sqrt{2}$ , 7L)      & \textbf{--2.9} (7.45 \AA)&  \textbf{1.1} (5.66 \AA) &     \\
                  &           &   ($2\times2\sqrt{2}$ , 9L)     & --6.8 (7.45 \AA)&     ---         &     \\\hline
    (110) Adatom  &    5      &   ($2\times\sqrt{2}$ , 5L)      & \textbf{--4.8} (7.45 \AA)& \textbf{11.9} (6.91 \AA) &  4.3\\
                  &           &   ($2\times\sqrt{2}$ , 7L)      &--14.2 (7.45 \AA)&     --6.5 (6.93 \AA)         &     \\
                  &           &   ($2\times2\sqrt{2}$ , 7L)     &--21.7 (7.45 \AA)&     ---         &     \\\hline
    (100) Adatom  &    4      &      ($2\times2$ , 5L)          & --6.5 (7.02 \AA)&--3.6 (6.51 \AA) &  7.3   \\
                  &           &      ($2\times2$ , 7L)          &   3.4 (7.02 \AA)& 38.7 (6.63 \AA) &     \\
                  &           &($\sqrt{2}\times\sqrt{2}$ , 5L)  &      ---        & 15.9 (9.35 \AA) &     \\
                  &           &($\sqrt{2}\times\sqrt{2}$ , 7L)  &   \textbf{2.6} (9.67 \AA)& \textbf{22.9} (9.28 \AA) &     \\
                  &           &($\sqrt{2}\times\sqrt{2}$ , 9L)  &      ---        & 31.9 (9.35 \AA) &     \\
                  &           &($\sqrt{2}\times\sqrt{2}$ , 11L) &       ---       & 25.1 (9.35 \AA) &     \\
    bulk-like (`random') & & ($2\times2\times2$) & --5.6 & --- &  \\
    bulk-like (adatom) & & ($2\times2\times2$) & 8.0 & --- &  \\
    bulk-like (free layer) & & ($2\times2\times2$) & --4.2 & --- &  \\
    \textit{bcc} bulk like (adatom) & & ($2\times2\times2$) & --9.6 & --- \\ \hline

    (111) Adatom  &    3      &($2\sqrt{2}\times2\sqrt{2}$ , 5L)&  \textbf{24.9} (8.11 \AA)& \textbf{31.9} (7.92 \AA) & 16.0 \\ \hline
    (111) Adatom  &    2      &    ($\sqrt{2}\times\sqrt{2}$ , 7L)    &      n.p.       & \textbf{33.4} (8.26 \AA) & --- \\
   \end{tabular}
  \end{ruledtabular}
 \end{center}
\end{table*}
Details on the cell are given in the format: (2D cell size ,
number of Ni layers). The Cd-Cd distance through the vacuum spacer
is given in parentheses besides the HFF values. The long hyphen
'---' is used to label cases that were not calculated. The NN=2
coordinated site has been achieved by placing the Cd atom in a
non-crystallographic adatom position on Ni(111). An `ideal
unrelaxed position' has therefore no meaning for Cd on this site,
which is indicated by the label `n.p.' (`not possible'). For the
fully coordinated bulk site, (NN=12) both a bulk and a slab
calculation are reported, the latter with the Cd placed in the
middle layer of the Ni slab. The experimental results of Potzger
\cite{potzger2002} are reported in the last column. We first
discuss the unrelaxed calculations, where all the atoms (including
Cd) sit at their ideal \textit{fcc} position. The calculated
values predict for the Cd HFF a change in sign for mid
coordination and large positive values for low coordination, in
agreement with the experimental assignments. Changing the size of
the slab, by adding Ni layers or by increasing the extension of
the cells in the surface plane, results into some scattered values
which lay, except for NN=5 and to a lesser extent also for NN=4,
within the aforementioned expected precision ($\pm$ 5T). For the
NN=5 case, i.e. adatom position at the Ni(110) surface, the HFF's
are found to attain large negative values for some of the
considered cells. We will come back to these puzzling results
later on.

As already discussed before, lattice relaxations are expected in
such open systems and might induce important changes to the HFF's.
In fact it has been shown recently that their inclusion improves
the agreement with the experimental data, in the case of 5sp and
6sp impurities in bcc bulk Fe. \cite{korhonen2000,cottenier2000}
Due to the larger atomic volume of Cd with respect to Ni, an
outward relaxation away from the surface is
expected for both terrace and adatom positions, for all low-index surfaces. Our calculations
show Cd displacements from ideal \textit{fcc} positions towards
the vacuum, that range from 0.60 to 0.90 \AA\ for the terrace atom
and from 0.09 to 0.26 \AA\ for the adatom, depending on the
surface orientation (Tab.~\ref{tab4}).
\begin{table}
 \begin{center}
  \caption{Cd perpendicular relaxations (in \AA) from ideal \textit{fcc}
   crystallographic site for the three low-index Ni surfaces. Everywhere Cd moves towards the vacuum. \label{tab4}}
  \begin{ruledtabular}
   \begin{tabular}{cccc}
     Site & (100) & (110) & (111) \\ \hline
    Adatom  & 0.20   & 0.26 & 0.09 \\
    Terrace & 0.60   & 0.90 & 0.84
   \end{tabular}
  \end{ruledtabular}
 \end{center}
\end{table}
Displacements for the terrace site are larger than for the adatom
site: the Cd atom strives for a Cd-Ni bond length of about 2.65
\AA\, that is somewhat larger than the ideal \textit{fcc} Ni-Ni
bond length of 2.48 \AA. Starting from an adatom position, this
can be realized with less displacement. Minor relaxations appear
also in the Ni atoms around the impurity. As evident from
Tab.~\ref{tab3}, the correction due to the relaxation is mostly
within 7~T (except for NN=4 with corrections up to 35~T, see
later). The correction is negative for the highest coordination
numbers (NN=9 and 12) and positive for all others.

\begin{figure}
 \begin{center}
  \includegraphics[width=6cm,angle=270]{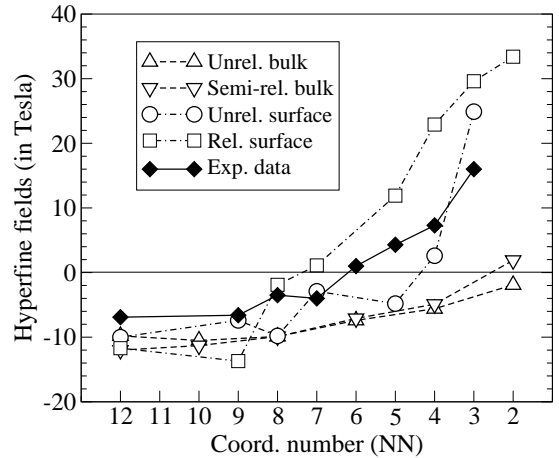}
  \caption{Calculated coordination number (NN) dependence of the
  Cd hyperfine field. Details are explained in the text. \label{fig2}}
 \end{center}
\end{figure}

For a better comparison, the theoretical results for the unrelaxed and
relaxed systems are plotted in Fig.~\ref{fig2}
together with the experimental results. When more than one value
exists in Tab.~\ref{tab3}, the ones relative to the cell with the
largest volume are selected, except for the more sensitive NN=4
and NN=5 cases where the values closer to the experiments are
chosen (this will be justified below). The chosen HFF's are given
in bold in Tab.~\ref{tab3}. As evident from Fig.~\ref{fig2}, upon
relaxation of the atoms in the cell the barycenter of the HFF
curve moves towards more positive values for NN $>$ 9, inducing a
sign change already for NN = 7 (Cd on a Ni(110) terrace site). The
largest changes are seen for NN=4 and NN=5, indicating once more a
high sensitivity of these environments. No clear overall
improvement is found when relaxations are included, and the
experimental data lays somehow between the relaxed and unrelaxed
theoretical curves.

\begin{figure}
 \begin{center}
  \includegraphics[width=6cm,angle=270]{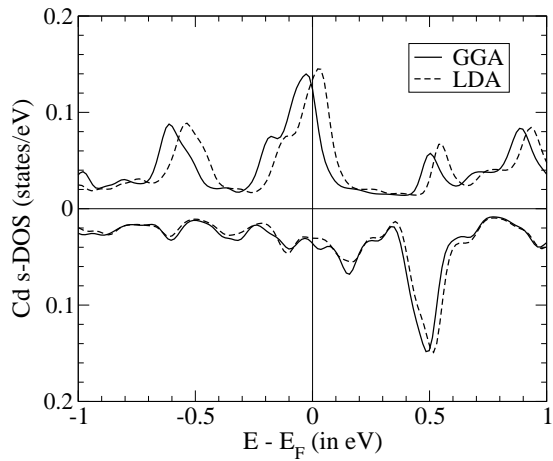}
  \caption{Partial s-DOS for relaxed Cd on Ni(100) calculated in the LDA and
  GGA approximations. \label{fig3}}
 \end{center}
\end{figure}

In order to investigate the reason for the large variations seen
for the NN=4 and NN=5 coordinated sites, it is fruitful to look at
the partial Density Of States (DOS) of Cd with s-symmetry (only
s-electrons contribute to the HFF). In Fig.~\ref{fig3}
the majority and minority s-DOS of a Cd adatom on Ni(100) is
shown in a small energy window in the vicinity of the Fermi
energy. Since it is not clear \textit{a~priori} whether the
problem arises due to the adopted approximation for the
exchange-correlation functional, LDA-calculations are presented as
well. The s-DOS shows several structures, among them a pronounced
peak which for the majority channel lays right at
E$_{\mathrm{F}}$ while it remains above E$_{\mathrm{F}}$ for the
minority channel. Small variations in the details of the cell
(number of layers, 2D cell size, relaxation or not, $\ldots$) or
in the computational method (LDA/GGA, APW+lo or KKR, $\ldots$)
will push this peak in the majority channel below or above
E$_{\mathrm{F}}$. Since only the majority spin is involved, one
has a net change of the s spin magnetic moment, and -- because
this is roughly proportional to the HFF -- also a change of the
HFF itself. For instance, in the two examples in Fig.~\ref{fig3},
the s spin moment and HFF are $2.02 \cdot 10^{-2}$ $\mu_B$ and
38.7~T for GGA, and $1.05 \cdot 10^{-2}\mu_B$ and 20.4~T for LDA.
We pinpoint here that the LDA does not generally solve this
sensitivity problem: in the NN=5 adatom unrelaxed case, for
instance, similar negative s moments and large negative HFF are
obtained both by LDA and GGA. As a general rule, whenever such a
peak is observed so close to E$_{\mathrm{F}}$, an enhanced
sensitivity of the HFF is foreseen, which adds to the inner
precision of the calculations. For NN=4 and NN=5, we do observe
such a peak in the s-DOS, for the other environments this is not
the case. This explains the instability and wide scattering of the
HFF's for those two coordination numbers in Tab.~\ref{tab3} (note
that this instability does not affect the calculations for
$V_{\mathrm{zz}}$, which are ruled by p- rather than s-electrons).
For impurity elements other than Cd, it might be that such
sensitivity does not show up at all, or -- if it does -- it might
do so for other values of the coordination number, depending on
the details of the s-DOS close to $E_F$. The experimental values
in Tab.~\ref{tab3} are up to 20~T below the calculated values for
NN$\ge$4. This indicates that in nature the majority s-peak is
above E$_{\mathrm{F}}$, while in our calculations it is below.

We now turn to the experimental parabolic NN-counting rule from
Fig.~\ref{fig2}. Both our relaxed and unrelaxed calculations show
roughly the same trend (Fig.~\ref{fig2}), and agree with the
experimental trend. But how absolutely does this `parabolic rule'
holds? Is it really true -- as concluded by Potzger et al.\
\cite{potzger2002} -- that knowledge of the NN coordination is
enough to predict the HFF on Cd in Ni-environments? In order to
answer this question, we exploit the advantage of \textit{ab
initio} calculations that one is not restricted to environments
that necessarily have to exist in nature. We can easily create
artificial Cd-in-Ni-environments with an arbitrary number of
nearest neighbors. That allows us to test the NN-counting rule in
more situations than are experimentally accessible. The
environments we created are 2x2x2 supercells for Cd in Ni, with a
given amount of nearest neighbor Ni atoms removed (such that
vacancies remain). This removal was done pair by pair, with the
requirement that the remaining cell still has inversion symmetry
(out of the several possibilities for every NN, we calculated only
one). This requirement makes such environments essentially
different from the corresponding surface environments with the
same NN, because the latter inevitably do not have inversion
symmetry. In a first series, we start from the relaxed Cd-in-Ni
supercell (labelled as `semi-relaxed',  because after removal of
the Ni atoms no further relaxation is done), in a second series
all atoms are at the ideal Ni bulk positions. The results of both
series are almost identical, and are reported in Fig.~\ref{fig2}.
Clearly, the same trend as in experiment and as in the surface
calculations is present in the artificial bulk calculations.  This
proves that there is a basic truth in the NN-counting rule. On the
other hand, the large difference between the bulk and surface
calculations -- especially for the low-coordination environments
-- is a clear sign that contributions from higher coordination
shells are not negligible.

If the requirement of inversion symmetry is removed, we can bridge
the gap between these bulk-like cases and the surface slabs by
calculating bulk-like cells where the first coordination shell is
exactly the same as on a specific surface site. We did this for
NN=8 and NN=4 (Tab.~\ref{tab3}, the environment labelled as
`random' is the one with inversion symmetry). For NN=8, we could
simulate in this way an environment that has exactly the same
NN-coordination as the (100) terrace site. If the NN-counting rule
were absolutely valid, we would find exactly the same HFF in the
random and terrace-like bulk case. The results are indeed quite
close: --9.9~T and --8.2~T, which are values that are also not far
from the slab calculation (--8.3~T). An even more daring test of
the counting rule, is to put Cd at a substitutional site of an
unrelaxed hypothetical \textit{bcc} Ni 2$\times$2$\times$2
supercell (16 atoms), with a lattice constant chosen such that the
Ni-Ni distance (and hence also the unrelaxed Cd-Ni distance) is
the same as in the \textit{fcc} case. Even in this very different
kind of environment, the calculated HFF of --8.9~T follows the
simple counting rule.  All this is different for NN=4. Apart from
the random environment (--5.6~T), we tested a configuration that
is identical to the \textit{fcc} (100) adatom case (8.0~T), a
configuration with all 4 Ni plus Cd in the same plane (a free
layer, --4.2~T), and in a \textit{bcc} cell a configuration that
is identical to a \textit{bcc} (100) adatom (--9.6~T). These 4
numbers prove that -- even in \textit{fcc}-based environments only
-- the exact spatial configuration of the Ni neighbors in the
first coordination shell can be important, leading to differences
of more than 13~T. It is not surprising to find this effect for
NN=4 rather than NN=8, the former being identified before as a
sensitive case. From this analysis we conclude that although the
NN-counting rule certainly indicates a trend, there can be
substantial deviations from it for specific environments. In such
sensitive environments, the spatial arrangement of the neighbors
is important as well. There is probably some luck involved that
for Cd in Ni the behavior in nature is so smooth as experimentally
observed, and there is no fundamental reason why the data could
not have been considerably more scattered around the parabolic
trend.


\subsubsection{HFF's of the 5sp series}
\label{results2b}

\begin{figure}
 \begin{center}
  \includegraphics[width=5.7cm,angle=270]{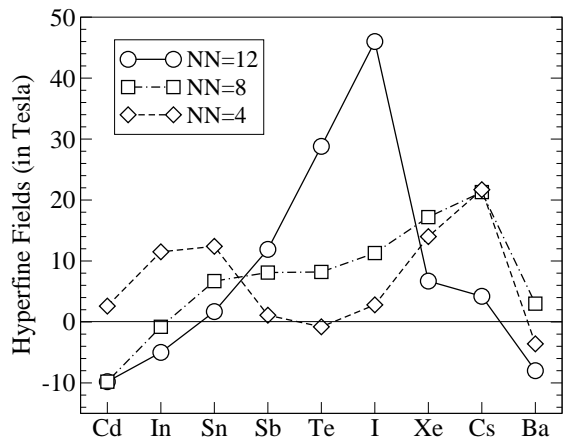}
  \caption{Hyperfine fields of the whole 5sp series (Cd $\rightarrow$Ba
  in bulk Ni (circle, NN=12), at a terrace position (squares, NN=8)
  and at an adatom position (diamonds, NN=4) on the Ni(100) surface. \label{fig4}}
 \end{center}
\end{figure}

As a last part of this study we now present a survey for the 5sp
impurities from Cd to Ba in Ni-environments, to see if and how the
NN-counting rule can be extended to other impurities. We use the
strategy applied by Mavropoulos et al.\ \cite{mavropoulos1998} for
4sp impurities on Fe and Ni surfaces, and very recently and
independently from this work also for 5sp impurities on Ni
surfaces. \cite{Mavropoulos2003} We take the bulk environment
(NN=12) and the terrace (NN=8) and adatom (NN=4) environments for
the (100) surface, and calculate the HFF for the 9 elements from
Cd to Ba in those environments. The cell sizes chosen for those
calculations are the ones labelled as in Tab.~\ref{tab3} as ($2 \times 2 \times 2$),
($2\times2$ , 5L) and ($2\times2$ , 7L), respectively. Because we
are looking again for gross trends now and not for fine details,
relaxations were not included. The results for the 5sp HFF's are
displayed in Fig.~\ref{fig4}.
Exactly the same behavior as Mavropoulos et al.\ observed for 4sp
and 5sp impurities is seen here, which mutually supports the
validity of the very different computational methods that were
used. In the bulk environment, the HFF starts at about --10~T for
Cd, strongly increases with increasing atomic number Z, and
reaches a maximum near the middle of the series: $\simeq$46~T for
I. Then it decreases again, and at the end of the series turns
back to values close to --10 T. When the coordination number is
reduced, the main peak of the HFF curve moves to heavier elements
and an additional structure -- that for NN=8 is more a broad
shoulder than a peak -- appears at the beginning of the series.
For NN=4 two clear structures are evident, and the HFF increases
and decreases (with less intense variations than for the bulk)
twice in the course of the 5sp series.

\begin{figure}
 \begin{center}
  \includegraphics[width=9cm,angle=0]{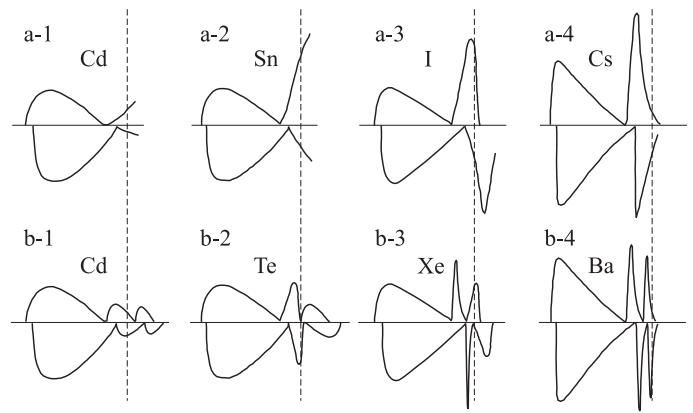}
  \caption{Cartoons for the majority and minority partial s-DOS of
  (a) a 5sp impurity in bulk Ni and (b) a 5sp impurity at a Ni surface. The
  vertical line indicates the Fermi energy, the name of the elements
  indicates for which element a particular picture is representative.
  This picture is inspired by Ref.~\onlinecite{akai1984}. \label{fig5}}
 \end{center}
\end{figure}

The microscopic origin of Fig.~\ref{fig4} can be understood by a
slight extension of arguments given by Mavropoulos et al.
\cite{mavropoulos1998} for 4sp impurities. Later on, we will then
derive (and test) generalized NN-counting rules for all 5sp
impurities in Ni environments from them. For a systematic
explanation, let us go back to the origin of hyperfine fields in
ferromagnets (see Ref. \onlinecite{akai1984} for a detailed and
instructive review). As the hyperfine field in our cases is
dominated by the Fermi contact contribution, we have to care about
the details of the bond between the 5s states and its environment
(here Ni-3d). It has been known for a long time
\cite{terakura1971,terakura1977} that in the case of such an s-d
bond the local s-DOS of the impurity shows a characteristic
depression a few eV below the fermi energy: the `antiresonance
dip' (AR). The position of the AR is mainly determined by the host
material (Ni), and not by the impurity. The states below the AR
are bonding states, the states above are antibonding. The up and
down states are exchange split, such that at first sight one
expects an excess of s-up over s-down, resulting in a positive
hyperfine field. Due to a different s-d hybridization for up and
down electrons, however, the number of s-up below AR will be
diminished, while the number of s-down will be enhanced. Above AR,
the situation is opposite. \cite{daniel1963} The final result is
that the impurity s-moment (and hyperfine field) will be negative
in the beginning of the sp-series, where the effect of the bonding
states is dominant (Fig.~\ref{fig5}-a-1).
In the second half of the series, also the antibonding states will
get filled, and because they have to be squeezed between AR and
the fermi energy, they have to develop a sharp peak in the DOS.
The exchange splitting of this peak is responsible for the large
positive HFF at the end of the sp-series (Fig.~\ref{fig5}-a-2/3),
which quickly drops to small and negative values again if also the
down antibonding states are below the Fermi energy
(Fig.~\ref{fig5}-a-4). Mavropoulos et al.\ have shown by group
theoretical arguments that in the case of reduced point group
symmetry for the impurity (as on surfaces), the antibonding part
of the impurity s-DOS is split in two parts. This splitting is
more pronounced if the impurity is in a more non-bulk-like
environments, i.e.\ it is more pronounced for NN=4 then for NN=8.
Let us take the NN=4 case with a clear splitting. When going from
Cd to Ba, we evolve through the different stages of
Fig.~\ref{fig5}-b, which explains the double-peak structure of the
HFF for NN=4 in Fig.~\ref{fig4}.

\begin{figure}
 \begin{center}
  \includegraphics[width=7cm,angle=0]{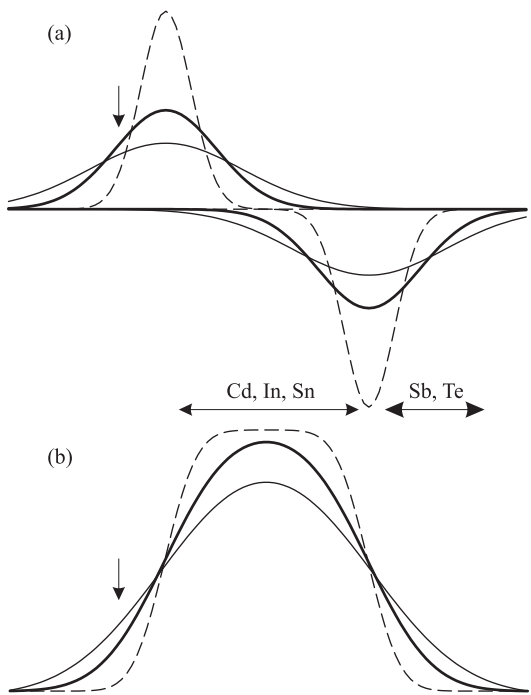}
  \caption{(a) The first of the two antibonding s-peaks for majority and
  minority spin. (b) The s-moment derived from (a) by subtracting --
  at a particular energy -- the integral of the minority s-DOS up to that
  energy from the integral of the majority s-DOS up to that energy.
  In both (a) and (b) 3 typical cases are drawn: high coordination =
  broad band width (thin full line), medium coordination = medium band width
  (thick full line) and low coordination = small band width (dashed line).
  The horizontal arrows indicate the region where the Fermi energy falls
  for the indicated elements (first half of the 5sp-series).
  This story is can be repeated with the second of the two antibonding
  peaks, starting from I for which the Fermi energy falls at the place indicated by
  the vertical arrow. \label{fig6}}
 \end{center}
\end{figure}

An aspect of Fig.~\ref{fig4} that has not been discussed by
Mavropoulos et al., is the physical origin of the coordination
number dependence of the HFF for a particular element: why is
e.g.\ for In the NN=4 HFF the larger one and the NN=12 the smaller
one, while this is reversed for e.g.\ Te? This we will explain by
the cartoon in Fig.~\ref{fig6}.
The upper part of Fig.~\ref{fig6} schematically shows the first of
the two antibonding s-peaks of Fig.~\ref{fig5}-b, for any
particular impurity. If one lowers the coordination number of the
impurity, the band-width of these peaks will decrease -- an
obvious fact, which we clearly observe in our calculations.
Fig.~\ref{fig6}-a shows the same situation for 3 typical
band-widths: large (NN=12), medium (NN=8) and small (NN=4). The
bottom part of Fig.~\ref{fig6} shows the s spin moment derived
from Fig.~\ref{fig6}-a as a function of energy (found by
subtracting the integral of the down-peak from the integral of the
up-peak, where the integrals are made up to the energy under
consideration). Everything now depends on where the Fermi energy
lies in Fig.~\ref{fig6}. If it falls in the region indicated by
Cd-In-Sn, Fig.~\ref{fig6}-a corresponds to Fig.~\ref{fig5}-b-1. At
the corresponding energy in Fig.~\ref{fig6}-b, the s-moment (HFF)
for the small band-width (NN=4) is larger then for the medium
band-width (NN=8), which in turn is larger than for the large
band-width (NN=12). If the Fermi energy falls in the region
indicated by Sb-Te, Fig.~\ref{fig6}-a corresponds to
Fig.~\ref{fig5}-b-2 and the sequence of s-moments is reversed.
After Te, this story repeats for the second antibonding peak (for
I, the Fermi energy will be at the position marked by the vertical
arrow, but of course in the second series of peaks), but will be
increasingly less clear due to the presence of the 6s states that
start to manifest themselves around the Fermi energy. That is the
reason for the more chaotic evolution for Cs and Ba. Of course,
Fig.~\ref{fig6} is a cartoon only, and its conclusions should not
be taken too literally: the real DOS are not Gaussians as used in
the cartoon, and therefore the details of the hyperfine field
evolution might be different. Nevertheless, it captures the basic
mechanism. Summarizing, we conclude that the physical mechanism
behind Fig.~\ref{fig4} can be understood from a combination of
three basic features: i)~the double peak structure of the
antibonding peaks, ii)~the decrease in the band width -- and hence
the increase of peak height -- upon reduction of the coordination
number, and iii)~the position of the Fermi energy with respect to
the peaks.

\begin{figure}
 \begin{center}
  \includegraphics[width=7cm,angle=270]{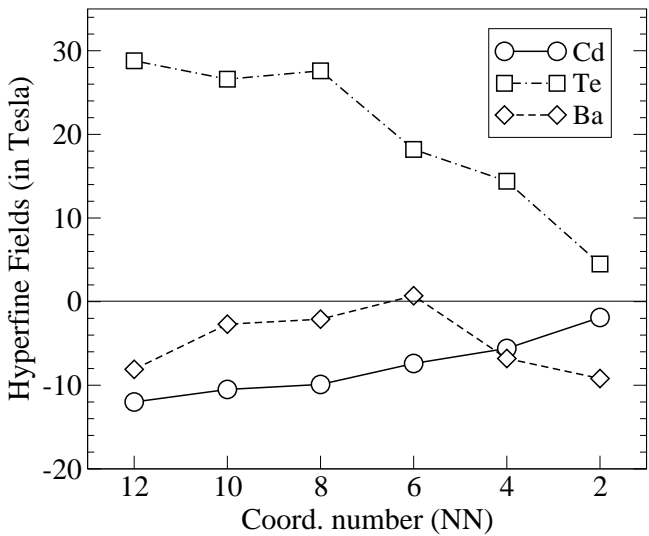}
  \caption{Coordination dependence of the HFF for selected, i.e. Cd,Te and Ba,
  5sp elements as obtained by bulk calculations (see text for details). \label{fig7}}
 \end{center}
\end{figure}

We now take Fig.~\ref{fig4} as a source of inspiration to extend
the parabolic NN-counting rule proposed by Potzger et al.\ (Ref.
\onlinecite{potzger2002}) for 5sp impurities other than Cd. It can
be seen from Fig.~\ref{fig4} that the Cd-HFF for bulk and NN=8 is
almost the same and negative, while the value for NN=4 is small
and positive: this is the parabolic behavior seen in experiment.
In the same way, we can then deduce that for In and Sn as
impurities, the HFF should monotonically rise from NN=12 to NN=4
(it is interesting to note that for the experimentally `easily'
accessible M\"{o}ssbauer probe $^{119}$Sn, Fig.~\ref{fig4}
suggests a linear behavior). Between Sn and Sb all lines cross,
such that for Sb to I the HFF monotonically decreases from NN=12
to NN=4. For Xe to Ba, there is non-monotonic behavior instead. We
have checked this deduction by calculating the artificial
bulk-like (non-relaxed) environments as discussed before, but now
for Te and Ba as impurities (both are taken as a representative
for the region of monotonic decrease and the non-monotonic region,
just as Cd is a representative of the region of monotonic
increase). In these bulk-like cells, we can more easily create
environments with NN different from 4, 8 and 12. The results are
given in Fig.~ \ref{fig7}.
The general trends shown in Fig.~\ref{fig5} are the same as the
ones which could be inferred from the surface calculations of
Fig.~\ref{fig4}: monotonic (parabolic) increase for Cd, monotonic
decrease for Te and non-monotonic behavior for Ba. In the light of
these results we therefore conclude that each of the 5sp
impurities in Ni has its own typical coordination number counting
rule.


\section{Conclusions}
\label{conclusions}

We have undertaken an extensive confrontation between state-of-the-art
\textit{ab initio} calculations and a high-quality data set -- experimentally
collected during the past 20 years -- of electric-field gradients and
magnetic hyperfine fields of Cd at magnetic and non-magnetic metallic fcc
surfaces. We conclude that the experimental practice for (100) and (111)
surfaces of assigning a large V$_{\mathrm{zz}}$ to terrace sites and a low
V$_{\mathrm{zz}}$ to adatom sites is correct. However, this `discrete' rule
is nothing more than a manifestation of a continuous evolution of the EFG
components as a function of coordination number, an evolution for which we
have pointed out the physical mechanism. With this insight, the behavior of
the (110) surface is not exceptional at all. The experimentally suggested
parabolic-like coordination number dependence for the HFF of Cd at Ni
surfaces is confirmed as being a reliable trend, but we warn that it is just
a trend and not a rigorous rule: sensitive environments exist, for which the
spatial arrangement of the Ni neighbors considerably influences the HFF. We
have explained in detail the physical mechanism behind the HFF for all 5sp
impurities at Ni surfaces, by combining knowledge from the literature and new
insight. In particular we have generalized the parabolic NN-counting rule for
Cd to other 5sp impurities, showing that each impurity has its own typical
rule, and explaining why this is so. We hope to have demonstrated that
\textit{ab initio} calculations can greatly enhance the physical insight in
an experimentally complex problem.


\begin{acknowledgments}
One of the authors (V.B.) acknowledges fruitful discussion with
Dr.~K.~Potzger. Part of the calculations were performed on
computer facilities granted by an INFM project \textit{Iniziativa
Trasversale Calcolo Parallelo} at the CINECA supercomputing
center. Another part of computational work was performed on
computers in Leuven, in the frame of project G.0239.03 of the
\textit{Fonds voor Wetenschappelijk Onderzoek - Vlaanderen} (FWO).
The authors are indebted to L.~Verwilst and J.~Knudts for building
and maintaining the pc-cluster in Leuven. M.C.\ thanks the
\emph{Onderzoeksfonds} of the K.U.Leuven for granting a
post-doctoral fellowship (F/02/010). Knowledge, help and advice
from the active WIEN2k user community are warmly acknowledged.
\end{acknowledgments}


\end{document}